\documentclass[aps,prd,twocolumn,showpacs,preprintnumbers,amsmath,amssymb,groupedaddress]{revtex4}
\usepackage{graphicx}
\usepackage{txfonts}

\begin{document}


\title{A Simple Optical Analysis of Gravitational Lensing}


\author{Xing-Hao Ye}
\email[Electronic address:]{yxhow@163.com}
\author{Qiang Lin}
\email[Electronic address:]{qlin@zju.edu.cn}

\affiliation{Department of Physics, Zhejiang University, Hangzhou
310027, China }


\date{\today}

\begin{abstract}
We analyzed the influence of static gravitational field on the
vacuum and proposed the concept of inhomogeneous vacuum. According
to the corresponding Fermat's principle in the general relativity,
we derived a graded refractive index of vacuum in a static
gravitational field. We found that the light deflection in a
gravitational field can be calculated correctly with the use of this
refractive index and therefore the gravitational lensing can be
treated conveniently with the optical method. For illustration, we
simulated the imaging of gravitational lensing, figured out the time
delay between the two images and calculated the lens mass in a
conventional optical way.
\end{abstract}


\maketitle


\section{Introduction}
\label{}

Gravitational lensing is an effect predicted by general relativity
\cite{Mollerach2002,Wambsganss1998,Ohanian1994}. It has been now a
powerful tool for the study of the so intractable problems in
astrophysics and cosmology such as the value of Hubble constant
\cite{Kundic1997}, the physics of quasars \cite{Solomon2003}, the
mass and mass distribution of a galaxy or a galaxy
cluster\cite{Wambsganss1998}, the large scale structure of the
universe \cite{Massey2007,Springel2006}, the existence of dark
matter \cite{Freeman2003,Inada2003,Wittman2000}, the nature of dark
energy \cite{Bennett2006,Bennett2005} and so on. Since a curved
spacetime is involved in relativistic treatment of the gravitational
lensing, inconvenience and intricacy are inevitable. So it is
valuable to seek for an alternative method to handle problems of
gravitational lensing in a simpler way.

In the framework of curved spacetime, the velocity of light in
vacuum is regarded as a constant $c$, which means that vacuum is
``homogeneous'' and ``isotropic'', i.e., vacuum does not differ from
place to place, and its refractive index is always 1. However, the
recent theoretical and experimental progresses demonstrate that such
concept of vacuum turns out to be inappropriate when there are
matters or fields within finite distance. For example, the vacuum
inside a microcavity is modified due to the existence of the cavity
mirrors, which will alter the zero-point energy inside the cavity
and cause an attractive force between the two mirrors known as
Casimir effect \cite{rf-Gies2006,rf-Emig2006}, which has been
verified experimentally \cite{rf-Lamoreaux1997,rf-Chan2001}. A
second example is that, under the influence of electromagnetic
field, vacuum can be polarized, which has led to astonishingly
precise agreement between predicted and observed values of the
electron magnetic moment and Lamb shift, and may influence the
motion of photons \cite{rf-Ahmadi2006}. Dupays \emph{et al}.
\cite{rf-Dupays2005} studied the propagation of light in the
neighborhood of magnetized neutron stars. They pointed out that the
light emitted by background astronomical objects will be deviated
due to the optical properties of quantum vacuum in the presence of a
magnetic field. Also in \cite{rf-Rikken2003}, Rikken and Rizzo
considered the anisotropy of the optical properties of the vacuum
when a static magnetic field $\textbf{B}_0$ and a static electric
field $\textbf{E}_0$ are simultaneously applied perpendicular to the
direction of light propagation. They predicted that magnetoelectric
birefringence will occur in vacuum under such conditions. They also
demonstrated that the propagation of light in vacuum becomes
anisotropic with the anisotropy in the refractive index being
proportional to ${\textbf{B}_0}\times{\textbf{E}_0}$.

The facts that the propagation of light in vacuum can be modified by
applying electromagnetic fields to the vacuum implies that the
vacuum is actually a special kind of optical medium
\cite{rf-Ahmadi2006,rf-Dupays2005}. This is similar to the Kerr
electro-optic effect and the Faraday magneto-optic effect in
nonlinear dielectric medium. This similarity between the vacuum and
the dielectric medium implies that vacuum must also have its inner
structure, which could be influenced by matter or fields as well.
Actually, the structure of quantum vacuum has already been
investigated in quite a number of papers
\cite{rf-Armoni2005,rf-Barroso2006,rf-Dienes2005}.

So, if we introduce an inhomogeneous vacuum within the framework of
flat spacetime instead of a curved spacetime with a homogeneous
vacuum, we will then find an equivalent but simpler method, i.e., an
optical method, to treat the problem of gravitational lensing. It is
just what we will do in the paper below.

\section{The vacuum refractive index in a gravitational field}
\label{} In order to seek for an optical way to the treatment of
gravitational lensing, we will start from a formula representing the
corresponding Fermat's principle for the propagation of light in a
static gravitational field  derived from the general relativity by
Landau and Lifshitz \cite{rf-Landau1975}:
\begin{eqnarray}
\delta \int {g_{00}}^{-1/2}dl=0,
\end{eqnarray}
where $dl$  is the local length element passed by light and measured
by the observer at position $r$ in the gravitational field, $r$ is
the distance from this element of light to the center of
gravitational matter $M$, $g_{00}$ is a component of the metric
tensor $g_{\mu\nu}$, $g_{00}^{-1/2}dl$ corresponds to an element of
optical path length. ${g_{00}}^{-1/2}=dt/d\tau$, where $d\tau$
represents the time interval measured by the local observer for a
light ray passing through the length $dl$, while $dt$ is the
corresponding time measured by the observer at infinity. Eq.(1)
could then be rewritten as
\begin{eqnarray}
\delta \int {g_{00}}^{-1/2}dl&=&\delta \int \frac{dt}{d\tau} dl\nonumber\\
&=&\delta \int \frac{dt}{d\tau}\frac{dl}{ds} ds\nonumber\\
&=&\delta \int n ds=0,
\end{eqnarray}
where $ds$ is the length element measured by the observer at
infinity, corresponding to the local length $dl$.

Eq.(2) shows that if we set the scale of length and time at infinity
as a standard scale for the whole gravitational space and time, the
propagation of light then satisfies the standard representation of
Fermat's principle, with the space --- actually the vacuum
influenced by the gravitational matter --- possessing a graded
refractive index given by
\begin{eqnarray}
n=\frac{dt}{d\tau} \frac{dl}{ds} = n_1  n_2,
\end{eqnarray}
where $n_1$ relates to the time transformation relation (i.e., the
time dilation effect) $dt/d\tau$ and $n_2$ relates to the space
transformation relation (i.e., the length contraction effect)
$dl/ds$.

\begin{figure}
\includegraphics{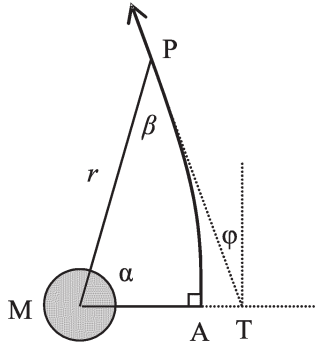}
\caption{\label{fig01} Light deflection caused by a graded
refractive index.}
\end{figure}

Now consider the light deflection caused by this graded refractive
index. In Fig.1, curve AP represents the light ray, $\beta$ is the
angle between the position vector $\textbf{r}$ and the tangent at
the point P on the ray, $\alpha$ is the angular displacement of the
vector $\textbf{r}$, $\varphi$ is the deflection angle of light .
For a gravitational matter with spherical symmetry, the refractive
index of the vacuum around will also be spherically symmetrical,
i.e., depends only on the distance $r$ for a given mass $M$. The
light propagation then satisfies the following relation
\cite{rf-Wolf1999}:
\begin{eqnarray}
n r \sin{\beta}=\textnormal{constant},
\end{eqnarray}
or
\begin{eqnarray}
n r \sin{\beta}=n_0 r_0,
\end{eqnarray}
where $r_0$ and $n_0$ represent the radius and refractive index at
the nearest point A respectively.

Since
\begin{eqnarray}
\tan{\beta}=\frac{rd\alpha}{dr},
\end{eqnarray}
associating with Eq.(5) reaches
\begin{eqnarray}
d\alpha=\frac{dr}{r \sqrt{(\frac{nr}{n_0 r_0})^2-1}}.
\end{eqnarray}

While the solution given by the general relativity
\cite{Mollerach2002,Arceo2006} reads
\begin{eqnarray}
d\alpha=\frac{dR}{R
\sqrt{\left[(\frac{R}{R_0})^2(1-\frac{2GM}{R_0c^2})+\frac{2GM}{Rc^2}\right]-1}},
\end{eqnarray}
where $R$, differing from the radius $r$ in the framework of flat
spacetime, is the radial coordinate in the Schwarzschild metric.

For an ordinary gravitational system, the gravitational field is not
extremely strong, i.e., $GM/rc^2<<1$, then we have the following
relations satisfying Eqs.(7) and (8):
\begin{eqnarray}
R=re^{\frac{GM}{rc^2}};
\end{eqnarray}
\begin{eqnarray}
n=e^{\frac{2GM}{rc^2}}.
\end{eqnarray}

\begin{figure}
\centering
\includegraphics{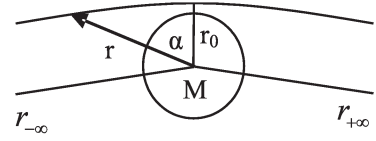}
\caption{\label{fig02} Light deflection in solar gravitational
field.}
\end{figure}

To verify the above expression of graded vacuum refractive index, we
consider a light ray passing by the Sun as shown in Fig.2. The total
angular displacement of the radius vector $\textbf{r}$ is
\begin{eqnarray}
\Delta \alpha=2 \int_{r_0}^{\infty} \frac{dr}{r \sqrt{(\frac{nr}{n_0
r_0})^2-1}},
\end{eqnarray}
where $r_0$ represents the nearest distance to the center of the
Sun. Substituting Eq.(10) into Eq.(11) gives a solution of first
order approximation
\begin{eqnarray}
\Delta \alpha=\pi + \frac{4GM}{r_0 c^2}.
\end{eqnarray}

Then the total deflection angle of light caused by the solar
gravitational field is
\begin{eqnarray}
\Delta \varphi = \Delta \alpha - \pi = \frac{4GM}{r_0 c^2},
\end{eqnarray}
which is fully consistent with that given by the general relativity
\cite{Ohanian1994,Weinberg1972} and the actual measurements
\cite{Fomaleont1976}, showing that the expression of the vacuum
refractive index we gave in Eq.(10) is workable.

\section{Optical Treatment of gravitational lensing}
\label{} The deflection of light caused by the gravitational field
of celestial bodies leads to the effect of gravitational lensing.
Formerly, this effect should be calculated complicatedly with the
general relativity \cite{Mollerach2002,Wambsganss1998,Ohanian1994}.
Once we have introduced the concept of graded vacuum refractive
index and obtained its relation with mass $M$ and position $r$, the
problem of gravitational lensing could then be treated easily with
the conventional optical method \cite{Ye2007}.

Considering a source $S$ and a lens $L$ of mass $M$, the light
emitted from $S$ is bent due to the gravitational field of the lens.
The bent light could be figured out through Eqs.(4) and (10).
Drawing the extension line of the light from the observer $O$, the
apparent (observed) position of the image could then be found out. A
computer simulated imaging is shown in Fig.3, where the upper image
$I_1$ and lower image $I_2$ are just drawn above and below the lens
$L$ for comparison between the three.

\begin{figure}
\centering
\includegraphics[totalheight=1.7in]{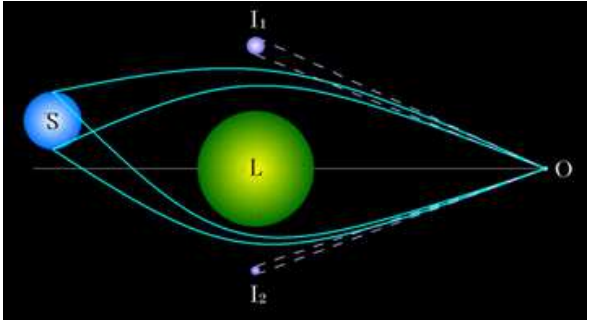}
\caption{\label{fig03}  Light path simulation of a gravitational
lensing.}
\end{figure}

In this way, the image shape could be figured out easily. Fig.4
shows a simulated result under the same conditions as that of Fig.3.
For comparison, the source (the small circle) is also drawn
proportionally in Fig.4, though it is blocked by the intervening
lens (the big circle). From the figure, we see that the images are
elongated tangentially. If source $S$ is located at axis $OL$, the
two images will be interconnected forming a ringlike image named
``Einstein ring'' (Fig.5). These results are consistent with the
known facts \cite{Mollerach2002,Wambsganss1998,Ohanian1994}.

\begin{figure}
\centering
\includegraphics[totalheight=1.7in]{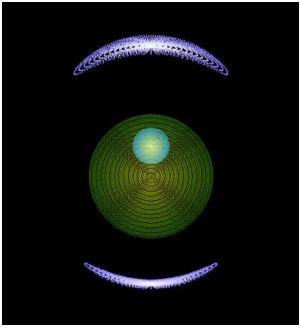}
\caption{\label{fig04}  Image shape simulation of a gravitational
lensing.}
\end{figure}

\begin{figure}
\centering
\includegraphics[totalheight=1.7in]{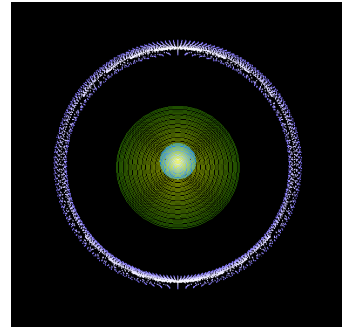}
\caption{\label{fig05} Einstein ring simulation.}
\end{figure}

\begin{figure}
\centering
\includegraphics[totalheight=1.2in]{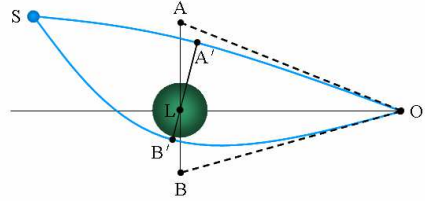}
\caption{\label{fig06} Time delay between the two images.}
\end{figure}

Another use of this optical method is to calculate the time delay
between the two images of a gravitational lensing.

According to Fig.1 and Eq.(5), we have
\begin{eqnarray}
dt = \frac{n ds}{c}=\frac{n dr/\cos \beta}{c} = \frac{n dr}{c
\sqrt{1-(\frac{n_0 r_0}{nr})^2}}.
\end{eqnarray}

Then, the time delay between the two images $A$ and $B$ in Fig.6 is
\begin{eqnarray}
\Delta t =t_B-t_A= \int_{SB'O} dt - \int_{SA'O} dt,
\end{eqnarray}
where $A'$ and $B'$ are the nearest points to the lens center for
the two light paths respectively.

Ordinarily, a gravitational lensing system satisfies
\begin{eqnarray}
r_S \ ( \textnormal{and: } r_O ) >> r_{A'} \ ( \textnormal{and: }
r_{B'} )
>> GM/c^2,
\end{eqnarray}
where $r_S$, $r_O$, $r_{A'}$, $r_{B'}$ represent the distances from
points $S$, $O$, $A'$, $B'$ to the lens center respectively.

Using Eqs.(10), (14), (15) and (16) we get
\begin{eqnarray}
\Delta t = \frac{4GM}{c^3} \ln{\frac{r_{A'}}{r_{B'}}}.
\end{eqnarray}

For example, in the case of the first observed gravitational lensing
--- the binary quasar Q0957+561, the measured time delay is $\Delta t =417 \pm 3 \  \textnormal
{days}$ \cite{Kundic1997,Glensdata2007}.

This time delay could be used to estimate the mass of the lens
galaxy:
\begin{eqnarray}
M = \frac {c^3 \Delta t}{4G \ln{\frac{r_{A'}}{r_{B'}}} }\approx
\frac {c^3 \Delta t}{4G \ln{\frac{\beta_{AOL}}{\beta_{BOL}}} }.
\end{eqnarray}

In the case of Q0957+561, $\beta_{AOL} / \beta_{BOL} \approx 5.5$,
thus we get $M \approx 2 \times 10^{42} \textnormal{kg}$. Since the
mass of our Milky Way galaxy is about $3.6 \times 10^{41}
\textnormal{kg}$, the mass we figured out here is an appropriate
estimation. Considering that the lens galaxy has a mass distribution
and image B is located among this distribution, the actual value of
the lens mass will be a little larger.

The result we obtained above can be further used to estimate the
Hubble constant. A simplified calculation for the cases of
Q0957+561, HE2149-2745 and RXJ1131-1231 \cite{Glensdata2007} gives
an average value $H_0 \approx 70\ \textnormal{km} / \textnormal{s}
\cdot \textnormal{Mpc}$.

\section{Conclusions}

We have proposed the concept of inhomogeneous vacuum with graded
refractive index based on the analysis of the influence of static
gravitational field on the vacuum. we derived the expression of this
refractive index through the general relativity and the optical
principle, and verified it in the effect of light deflection caused
by a gravitational field. By using this expression, we investigated
the gravitational lensing in a conventional optical way and showed
some computer simulations for the imaging of gravitational lensing.
Also in this way, we figured out the time delay between the two
images of a gravitational lensing and gave a simple calculation of
the lens mass, which could be further used to find out the value of
Hubble constant. The results indicate that, the concept of
inhomogeneous vacuum is mathematically equivalent to the curved
spacetime in the general relativity. In addition to the equivalence
and convenience, the investigation of an inhomogeneous vacuum
(actually a quantum vacuum) instead of a curved spacetime also
promises an approach to the physical connection between the general
relativity and the quantum mechanics.


$\/$
\appendix{\textbf{Acknowledgments}}

We wish to acknowledge the supports from the National Key Project
for Fundamental Research (grant no. 2006CB921403), the National
Hi-tech project (grant no. 2006 AA06A204) and the Zhejiang
Provincial Qian-Jiang-Ren-Cai Project of China (grant no.
2006R10025).

\end{document}